\RequirePackage[2020-02-02]{latexrelease}
\documentclass[%
 reprint,
 amsmath,amssymb,
 aps,
]{revtex4-2}
\usepackage[utf8]{inputenc}
\usepackage{pgfplots}

\begin{document}
\preprint{APS/123-QED}


\title{Dimensionless parameter of structural ordering and excess entropy of metallic and tellurite  glasses}

\author{A.S. Makarov}
\affiliation{Department of General Physics, Voronezh State Pedagogical
University, Lenin St. 86, Voronezh 394043, Russia}
\author{J.C. Qiao}
\affiliation{School of Mechanics, Civil Engineering and Architecture, Northwestern Polytechnical University, Xi’an 710072, China}

\author{G.V. Afonin}
\affiliation{Department of General Physics, Voronezh State Pedagogical
University, Lenin St. 86, Voronezh 394043, Russia}
\author{R.A. Konchakov}
\affiliation{Department of General Physics, Voronezh State Pedagogical
University, Lenin St. 86, Voronezh 394043, Russia}
\author{A.N. Vasiliev} 
\affiliation{National University of Science and Technology MISiS, Moscow 119049, Russia}
\author{V.A. Khonik} 
\email{v.a.khonik@yandex.ru}
\affiliation{Department of General Physics, Voronezh State Pedagogical
University, Lenin St. 86, Voronezh 394043, Russia}
\author{N.P. Kobelev}
\affiliation{Institute for Solid Solids Physics RAS, Moscow district, Chernogolovka 142432, Russia}

\date{\today}

\begin{abstract}
 
Using a notion on the excess entropy of glass with respect to the counterpart crystal, we introduce a simple dimensionless order parameter $\xi$, which changes from $\xi \rightarrow 0$ to $\xi \rightarrow 1$. The former case corresponds to a strongly disordered liquid-like structure while the latter situation describes a highly ordered crystal-like glass.  This approach is applied to 13 metallic and 2 tellurite glasses. We found that $\xi$ is strongly sensitive to structural state and/or chemical composition. It can be also used for a  comparison of the order in glasses belonging to different classes and appear to  represent  a new way of structural analysis.

\end{abstract}

\maketitle




\paragraph*{Introduction.} Glasses constitute a disordered state of condensed matter, which is prone to spontaneous atomic rearrangements driven by a decrease of the Gibbs free energy and leading to an increase of the entropy. This large-scale structural evolution is generally termed as structural relaxation \cite{Varshneya2019}. Amongst different types of glasses (oxide, chalcogenide, fluorine, polymeric, etc), structural relaxation in metallic glasses (MGs) is most pronounced \citep{GreerMetGlas2014}. It has been reliably documented that structural relaxation affects most (if not all) of the physical properties of MGs \cite{WangProgMaterSci2012,ChengProMatSci2011}. It is quite evident that structural relaxation is intrinsically related to structural ordering. The latter, according to the  statistical physics \cite{Landau}, should be related to the entropy of glass. This simple constatation should provide a general thermodynamic way to study structural relaxation and structural ordering in non-crystalline systems. However, rather numerous studies utilizing thermodynamic approach were directed to the calculation of the excess enthalpy, entropy and Gibbs free energy of the supercooled liquid state with respect to the maternal (counterpart) crystal. The obtained results were used mostly to estimate the glass-forming ability of supercooled liquids and their crystallization kinetics (e.g. Refs \cite{GladeJAP2000,WildeMSE2004,LuPRL2003,JiJNCS2007,JiangJALCOM2020}). At that, the thermodynamic potentials of MGs with respect to the maternal crystalline state remained out of view, with the exception of recent works \cite{MakarovJPCM2021,MakarovJETPLett2022}, which are discussed below.  

Meanwhile, a general thermodynamic approach to the description of structural ordering of non-crystalline solids was suggested in the 80s of the past century by Nemilov  \cite{NemilovFHS1981,NemilovFHS1982}. He argued that the degree of structural nonequilibrium is generally determined by the ratio $\alpha = \Delta S_0/\Delta S_{melt}$, where $\Delta S_0$ is the excess entropy of glass with respect to the maternal crystal at temperatures $T\rightarrow 0$ K and $\Delta S_{melt}$ is the crystal melting entropy. Respectively, the degree of glass structural equilibrium is $\xi = 1-\alpha$. Therefore, if glass is completely disordered then $\Delta S_0\approx\Delta S_{melt}$ and the parameter of structural order $\xi=0$. For a completely ordered glass (which is essentially the maternal crystal) $\Delta S_0=0$ and the order parameter $\xi=1$. Using this approach, Nemilov calculated the order parameter $\xi$ for a large number of inorganic, molecular organic and polymeric glasses \cite{NemilovFHS1982,NemilovJNCS2009} and determined the relations of this thermodynamic approach with structural peculiarities of oxide glasses \cite{NemilovFHS1982_2,NemilovFHS1983}. We are unaware of any implementations of this approach since the last Nemilov's work on this topic \cite{NemilovJNCS2009}.   

Meanwhile, this approach requires the knowledge of the heat capacities of glass, supercooled liquid and maternal crystal in the whole temperature range from $T\rightarrow 0$ K up to the melting temperature $T_m$. For MGs, this information is most often unavailable  and this may be the reason for the absence of  corresponding studies in the literature. However, this approach can be simplified and made more convenient to use by the introduction of the excess entropy $\Delta S$ of glass with respect to the maternal crystal at a  temperature $T$ taken equal to room temperature (rt) or above it. The structural order parameter then becomes 

\begin{equation}
\xi(T)=1-\frac{\Delta S(T)}{\Delta S_{melt}}, \label{Xi}
\end{equation}
where $\Delta S_{melt}$ is the entropy increase upon heating from the solidus $T_{sol}$ to liquidus $T_{liq}$ temperature. The order parameter varies in the range $0<\xi<1$ upon changing the structure from the fully disordered (liquid-like) state with $\Delta S \rightarrow \Delta S_{melt}$ and $\xi \rightarrow 0$ towards the fully ordered state (crystal-like) characterized by $\Delta S \rightarrow 0$ and $\Delta S\rightarrow 1$. The present work is devoted to the application of this method for 13 metallic  and 2 tellurite glasses.

\paragraph*{Experimental procedure.} The  method of calorimetric determination of the excess entropy $\Delta S$ for MGs was recently suggested in Refs \cite{MakarovJPCM2021,MakarovJETPLett2022}.  If $W_{gl}$ and $W_{cr}$ are the heat flows coming from glass and maternal crystal in differential scanning calorimetry (DSC) measurements, respectively, then the excess entropy of glass with respect to the maternal crystal is

\begin{equation}
\Delta S(T)=\frac{1}{\dot{T}}\int_{T}^{T_{cr}} \frac{\Delta W(T)}{T}dT, \label{DeltaS}
\end{equation}   
where $\Delta W(T)=W_{gl}(T)-W_{cr}(T)$ is the differential heat flow, $\dot{T}$ is the heating rate and $T_{cr}$ is the temperature of the complete crystallization. If  current temperature $T=T_{cr}$ then $\Delta S=0$ and, therefore, the function $\Delta S (T)$  represents temperature dependence the excess entropy of glass with respect to the maternal crystal. 

\begin{figure}[t]
\center{\includegraphics[scale=0.7]{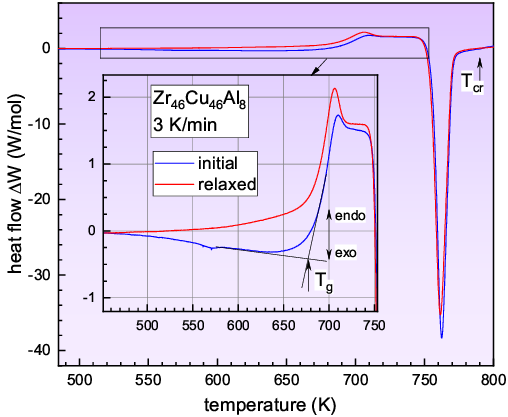}}
\caption[*]{\label{Fig1.eps}  Differential heat flow $\Delta W$  of glassy  Zr$_{46}$Cu$_{46}$Al$_{8}$ in the initial and relaxed states. The glass transition temperature $T_g$ and the temperature of the complete crystallization $T_{cr}$ are  indicated. The inset gives the glass transition and supercooled liquid regions on an enlarged scale.}
\end{figure} 

DSC measurements were performed using a Hitachi DSC 7020 instrument operating in high-purity (99.999$\%$) $N_2$ atmosphere at a rate of 3 K/min. Measurements were carried out on 13 x-ray  amorphous bulk MGs and  2 tellurite glasses produced by melt quenching and listed in Table 1. Every glass was tested as follows: \textit{i}) in the initial state up to  above $T_{cr}$ (sample 1), \textit{ii}) in the initial state up to the supercooled liquid state (i.e.  deep above above the glass transition temperature $T_g$)  and cooling back to room temperature at the same rate performing thus a relaxed state (sample 2 run 1), \textit{iii}) the same sample up to $T_{cr}$ (DSC on the relaxed sample 2, run 2) and \textit{iv}) the same (=crystallized) sample up to $\approx 900 $ K. This allowed to perform the above protocol with the reference DSC cell containing fully crystallized sample (prepared by the step \textit{i})) of approximately the same mass ($\approx 100-110$  mg).

\paragraph*{Results.} Figure \ref{Fig1.eps} gives $\Delta W$ traces of glassy Zr$_{46}$Cu$_{46}$Al$_{8}$ in the initial and relaxed states taken as an example. The initial state is characterized by clear exothermal reaction below $T_g$, which is a manifestation of structural relaxation. This reaction is absent in the relaxed state obtained by preheating up to 730 K (deep in the supercooled liquid region) and cooling back to room temperature at the same rate. Above $T_g$, a pronounced endothermal effect is observed evidencing the attainment of a supercooled liquid state. This effect is followed by a strong crystallization-induced exothermal reaction. Overall, both $\Delta W$ curves are quite similar to those obtained on other bulk MGs \cite{MakarovJPCM2021,MakarovJETPLett2022}.     

\begin{figure}[t]
\center{\includegraphics[scale=0.3]{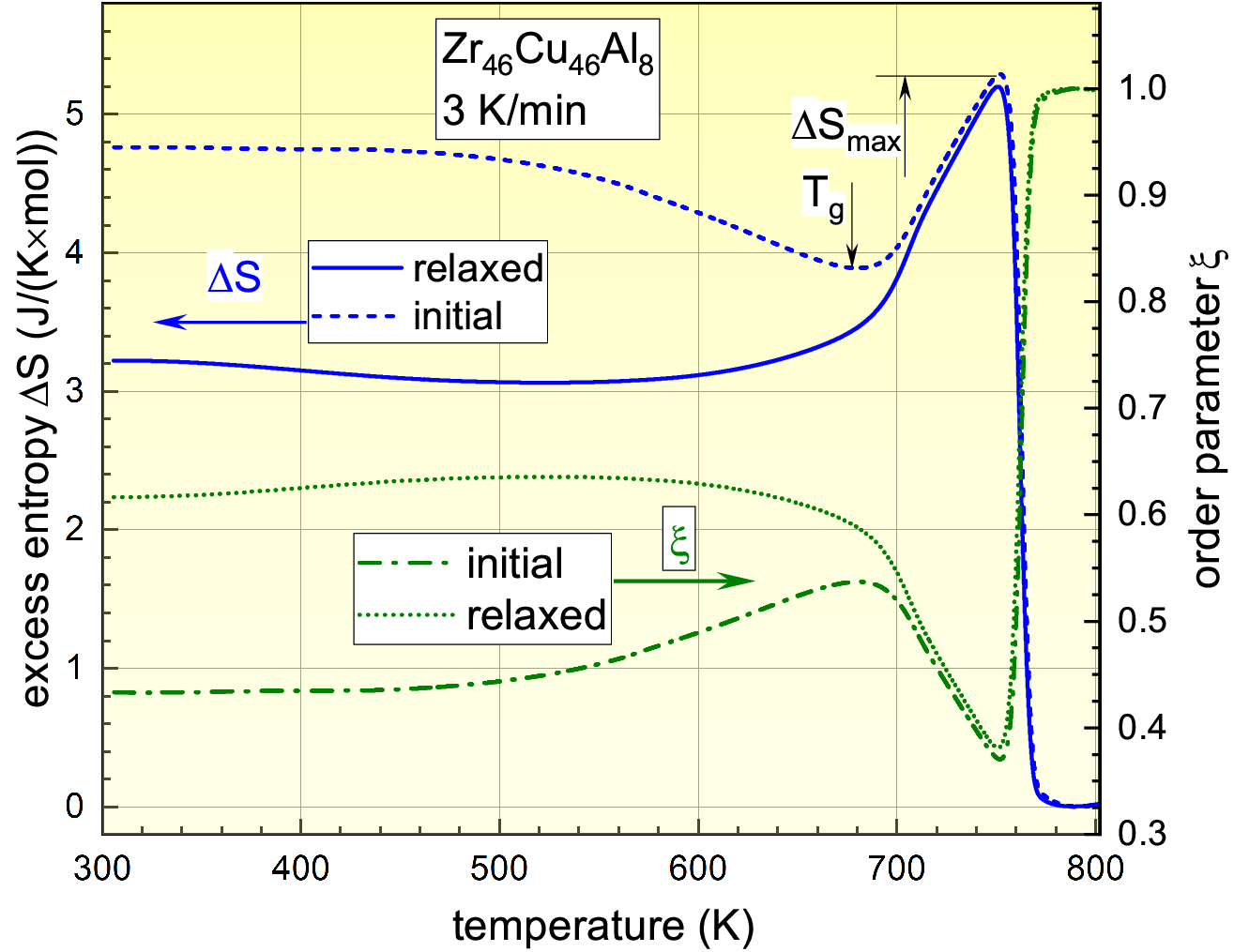}}
\caption[*]{\label{Fig2.eps}  Temperature dependences of the excess entropy $\Delta S$ and order parameter $\xi$ of glassy  Zr$_{46}$Cu$_{46}$Al$_{8}$ in the initial and relaxed states. }
\end{figure} 

The data in Fig.\ref{Fig1.eps} were used to calculate temperature dependences of the excess entropy $\Delta S$ using Eq.(\ref{DeltaS}), which are shown in Fig.\ref{Fig2.eps}. In the initial state near room temperature and up to $\approx 500$ K, $\Delta S\approx 4.8$ J/(K$\times$mol). Continued heating results in a decrease of this quantity to $\approx 3.9$ J/(K$\times$mol) just below $T_g$ because of ongoing structural relaxation (see also exothermal reaction in the same range in Fig.\ref{Fig1.eps}). Above $T_g$, one observes a rapid increase of the excess entropy up to its maximal value $\Delta S_{max}$, which is followed by a rapid $\Delta S$-drop to zero due to continued crystallization.  

One can now calculate the order parameter $\xi$ as a function of temperature using Eq.(\ref{Xi}). The order parameter $\xi(T)$ thus determined with the melting entropy $\Delta S_{melt}=8.4$ J/(K$\times$mol) (derived using the data presented in Ref.\cite{AfoninScrMater2019}) is shown in Fig.\ref{Fig2.eps}. In the initial state, $\xi\approx 0.43$ and remains constant up to $\approx 500$ K while structural relaxation above this temperature leads to a significant increase of this parameter so that $\xi=0.54$ at $T=T_g$, which corresponds to a substantial structural ordering. In the  supercooled liquid region above $T_g$, the order parameter rapidly falls down to $\xi=0.37$ (thermally activated structural disordering) and while subsequent crystallization leads to an increase of $\xi$ up to unity at $T=T_{cr}$ that means the attainment of the fully ordered (crystalline) state. Preannealing of glass by heating into the supercooled liquid region results in  a $\xi$-increase up to 0.61 at room temperature (relaxation-induced structural ordering). Upon heating of the relaxed sample, $\xi$ remains approximately constant up to $T_g$ while at higher temperatures $\xi(T)$-dependence repeats that of the sample in the initial state.        
           
One can conclude from Fig.\ref{Fig2.eps} that \textit{i})  changes of the $\xi$-parameter qualitatively agree with the expected variations of structure in different temperature regions (ordering and disordering below and above $T_g$, respectively) and sample's state (initial/relaxed) and \textit{ii}) this parameter is very sensitive  to these variations. In this regard, one can compare quite minor changes of the heat flow shown in Fig.\ref{Fig1.eps} with significant changes of the $\xi$-parameter in Fig.\ref{Fig2.eps}.   

\begin{table*}[t]
\caption{\label{tab:table1} Glasses under investigation and their melting entropy $\Delta S_{melt}$.  The latter quantity was derived  using the data presented in Ref. \cite{AfoninScrMater2019} and this work (except the compositions 9 and 13). Tellurite glasses 14 and 15 were tested only in strongly preannealed state \cite{Afonin ScrMater2020}.} 

\begin{tabular}{c|c|c}
\hline
\hline
No & Glass composition (at.\%) & Melting entropy   \\
   & & $\Delta S_{melt}$ (J/(K$\times$mol))     \\ 
\hline
\hline
1 & $Pd_{43.2}Cu_{28}Ni_{8.8}P_{20}$ &  7.3   \\ 

2 & $Pd_{40}Cu_{30}Ni_{10}P_{20}$    &  9.2  \\

3 & $Pd_{40}Ni_{40}P_{20}$           &   14.2   \\

4 & $Zr_{65}Al_{10}Ni_{10}Cu_{15}$   &   13.6   \\

5 & $Zr_{46}Cu_{45}Al_7Ti_2$       &   8.5   \\

6 & $Zr_{46}Cu_{46}Al_8$          &   8.4    \\

7 & $Zr_{47}Cu_{45}Al_7Fe_1$       &   8.6     \\

8 & $Zr_{55}Co_{25}Al_{20}$   &   7.4    \\

9 & $Zr_{52.5}Ti_5Cu_{17.9}Ni_{14.6}Al_{10}$ & 7.6 \cite{GladeJAP2000} \\

10 & $La_{55}Al_{35}Ni_{10}$        &   8.7    \\

11 & $La_{55}Al_{25}Co_{20}$         &  10.9    \\

12 & $Pt_{20}Pd_{20}Cu_{20}Ni_{20}P_{20}$ & 9.9  \\

13 & $Pt_{42.5}Cu_{27}Ni_{9.5}P_{21}$ & 11.3 \cite{NeuberActaMater2021}  \\

14 & $(TeO_2)_{45}(V_2O_5)_{55}$ & 27.1  \\

15 & $(TeO_2)_{70}(CuO)_{30}$ & 23.4  \\

\hline
\hline

\end{tabular}
\end{table*}

\begin{figure}[t]
\center{\includegraphics[scale=0.7]{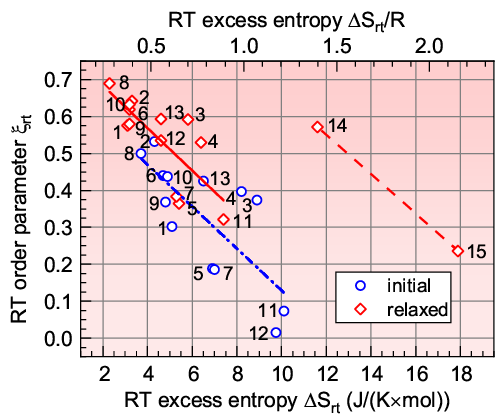}}
\caption[*]{\label{Fig3.eps}  Room-temperature order parameter $\xi_{rt}$ as a function of the excess entropy at room temperature $\Delta S_{rt}$ for the glasses in the initial and relaxed states. The numbers correspond to the  compositions given in Table 1. The solid lines give corresponding least square fits. }
\end{figure} 

Similar experiments were performed on all other glasses listed in Table 1. This list includes conventional bulk MGs, high-entropy MGs and two tellurite glasses  taken as representatives of another glass type and because their thermal behavior is quite similar to that of MGs \cite{Afonin ScrMater2020}. The melting entropy  was calculated as $\Delta S_{melt}=\frac{1}{\dot{T}}\int_{T_{sol}}^{T_{liq}} \frac{W_{gl}(T)}{T}dT$. Figure \ref{Fig3.eps} shows room-temperature order parameter $\xi_{rt}$ calculated using Eq.(\ref{Xi}) as a function of the excess entropy $\Delta S_{rt}$ at room temperature for the glasses in the initial and relaxed states. It is seen that an increase of $\Delta S_{rt}$ results in strong decrease of  the order parameter, as expected. At that, the derivative $d\xi_{rt}/d\Delta S_{rt}$ is the same for initial/relaxed/tellurite glasses. The entropy $\Delta S_{rt}$ of tellurite glasses 14 and 15 is several times larger compared with that of relaxed MGs but the order parameter is quite similar. 

The peculiarities of $\xi_{rt}(\Delta S_{rt})$-dependence can be summarized as follows: \textit{i}) a few MGs in the initial state at room temperature display structural order similar to that of a liquid (compositions 11,12);  
\textit{ii}) preliminary structural relaxation performed by heating into the supercooled liquid state results in a significant increase of the order parameter, as anticipated; \textit{iii}) glasses of close chemical compositions can have fairly different structural order (e.g. compare compositions 1 and 2; compositions 6 and 7);  \textit{iv}) compositionally very different glasses can have close $\xi$  (e.g. compositions 4 and 10) and \textit{v}) the excess entropy $\Delta S_{rt}$ of MGs in the initial and relaxed states varies in a wide range $0.2\leq R \leq 1.2$ ($R$ is the universal gaz constant). One can also notice that  $\Delta S_{rt}$-values on the average amount to $\approx 70$\% and  $\approx 50$\% of the melting entropy $\Delta S_{melt}$ for the initial and relaxed MGs, respectively. 

\begin{figure}[t]
\center{\includegraphics[scale=0.7]{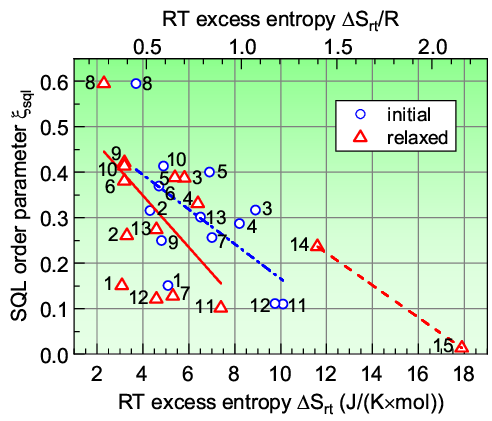}}
\caption[*]{\label{Fig4.eps}  Order parameter $\xi_{sql}$ calculated for the supercooled liquid state as a function of the excess entropy at room temperature $\Delta S_{rt}$.}
\end{figure} 

It is interesting to calculate the order parameter for the supercooled liquid (sql) state. For this, we used Eq.(\ref{Xi}) with the maximal excess entropy $\Delta S_{max}$ (as exemplified in Fig.\ref{Fig2.eps}), which corresponds to the supercooled liquid state just before crystallization onset. The excess entropy $\xi_{sql}$ thus determined is shown in Fig.\ref{Fig4.eps} as a function of the room-temperature excess entropy $\xi_{rt}$. It is seen that $\xi_{sql}$-values for every glass in the initial and relaxed states are relatively close suggesting the independence of the structural order in the supercooled liquid state on samples' thermal prehistory, in line with many previous observations (e.g. Ref.\cite{MakarovMetals2022}). At that, the derivative $d\xi_{sql}/d\Delta S_{rt}$, accounting for the scatter, is about the same for initial/relaxed/tellurite glasses. Of special note is that the data on tellurite glasses fall into the general trade.   

\begin{figure}[b]
\center{\includegraphics[scale=0.7]{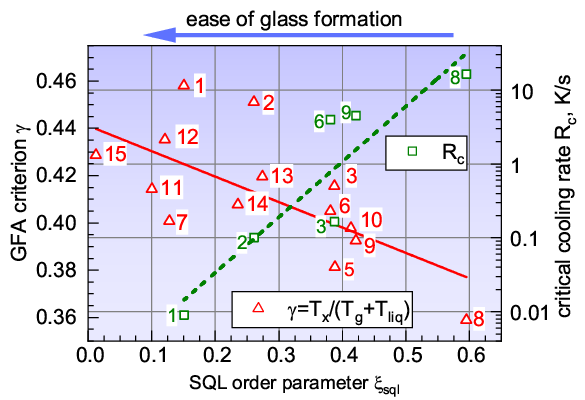}}
\caption[*]{\label{Fig5.eps} Critical cooling rate $R_c$ and glass formation criterion  $\gamma=T_{x}/(T_g+T_{liq})$ for the indicated glasses (Table 1) as a function of the order parameter $\xi_{sql}$ in the supercooled liquid state. The solid lines give corresponding mean-square fits. The $R_c$-data are taken from Refs \cite{ShenApplPhysLett2007,InoueMaterTransJIM1997} (glasses 1-3), \cite{LanApplPhysLett2014} (glass 6), \cite{MukherjeePRB2004} (glass 8) and \cite{XingJALCOM2004} (glass 9). }
\end{figure} 

It is clear that the structure of a supercooled liquid is responsible for the glass-forming ability (GFA), and, therefore, one can suppose that the order parameter $\xi_{sql}$ in this state should be  related to the GFA as well. This hypothesis is tested in Fig.\ref{Fig5.eps}, which gives the data on the melt critical cooling rates  available in the literature for the glasses under investigation and the  criterion $\gamma=T_{x}/(T_g+T_{liq})$ ($T_x$ is the crystallization onset temperature) widely known to provide a good GFA characterization \cite{LuPRL2003}. It is seen that $R_c$ strongly increases with  $\xi_{sql}$, which seems to be quite reasonable: the higher the structural order in the supercooled liquid state is the more difficult  the glass formation should be.  On the other hand, the critical cooling rate is known to decrease with the parameter $\gamma$ \cite{LuPRL2003} reflecting an increase of the GFA. According to Fig.\ref{Fig5.eps} this behavior originates from a decrease of structural order described by $\xi_{sql}$. This understanding also agrees with a fundamental work by Li et al. \cite{LiNatureMater2022}  who fabricated over five thousand alloys and found a strong correlation between the high GFA and a direct measure of glass disorder given by the width of X-ray diffraction structure factor. Thus, the GFA allows a qualitative understanding in terms of the structural order parameter $\xi_{sql}$ in the supercooled liquid state. 

\paragraph*{Concluding remarks.} The dimensionless order parameter $\xi$ changes in the range $0<\xi <1$ and  provides  useful qualitative and quantitative information on structural ordering in various glasses and its evolution upon  relaxation. This order parameter is quite sensitive to the glass state and its chemical composition. It shows that some MGs display structural order similar to that of the liquid at the liquidus point. Structural relaxation below $T_g$ results in a significant increase of the order parameter.  At that, glasses close in chemical composition can have fairly different  structural order while compositionally different glasses can display close order parameter. 

We also show that the structural order parameter $\xi_{sql}$ for the supercooled liquid state is related with the critical melt cooling rate $R_c$ and the glass formation parameter $\gamma$. An increase of the structural order given by $\xi_{sql}$ worsens the glass forming ability (i.e. increases $R_c$ and decreases $\gamma$).   

The data on tellurite glasses fall into a general trend characteristic of   metallic glasses. There are clear grounds to assume that the above approach is suitable for a comparison of structural order and its changes upon relaxation of glasses belonging to different classes. It also turns out that the information obtained with this approach cannot be obtained otherwise. At that, the approach is based on a simple and transparent idea given by Eq.(\ref{Xi}) while the excess entropy of glass with respect to the maternal crystal and the melting entropy  entering this equation can be relatively easily derived from calorimetric measurements.   

The work was supported by the Russian Science Foundation under the project No. 23-12-00162.

\end{document}